\documentclass[
preprint,
prb,
amsmath,amssymb,
aps,
]{revtex4-2}

\usepackage{graphicx}
\usepackage{dcolumn}
\usepackage{bm}

\usepackage{xcolor}
\usepackage[normalem]{ulem}

\graphicspath{{eps/}{png/}}

\begin{document}


\title{Momentum alignment and the
optical valley Hall effect in low-dimensional Dirac materials
}

\author{V. A. Saroka}
\affiliation{Physics and Astronomy, University of Exeter, Stocker Road, Exeter EX4 4QL, United Kingdom}
\affiliation{Institute for Nuclear Problems, Belarusian State University, Bobruiskaya 11, 220030 Minsk, Belarus
}

\author{R. R. Hartmann}
\affiliation{
Physics Department, De La Salle University,  2401 Taft Avenue,
0922 Manila, Philippines
}

\author{M. E. Portnoi}
\email{M.E.Portnoi@exeter.ac.uk}
\affiliation{Physics and Astronomy, University of Exeter, Stocker Road, Exeter EX4 4QL, United Kingdom}

\date{\today}

\begin{abstract}
We study the momentum alignment of photoexcited carriers and the optical control of valley population in gapless and gapped two-dimensional Dirac materials. The trigonal warping effect leads to the spatial separation of charge carriers belonging to different valleys upon linearly polarized high-frequency photoexcitation. Valley separation in gapped materials can be detected by measuring the degree of circular polarization of band-edge photoluminescence at different sides of the sample or light spot (optical valley Hall effect). We demonstrate that the celebrated Rashba effect, caused by substrate-induced system asymmetry, leads to a strong anisotropy in the low-energy part of the spectrum. This results in optical valley separation by a linearly polarized excitation at much lower frequencies compared to the high-energy trigonal warping regime. We also show that the momentum alignment phenomenon explains the giant enhancement of near-band-edge interband optical transitions in narrow-gap carbon nanotubes and graphene nanoribbons independent of the mechanism of the gap formation.
These enhanced transitions can be used in terahertz emitters based on low-dimensional Dirac materials.
\end{abstract}

\keywords{graphene, 2D materials, carbon nanotubes, graphene nanoribbons, valeytronics}


\maketitle

\section{Introduction}
One of the best-known properties of graphene is its universal optical absorption for a broad range of frequencies~\cite{Nair_Science_08}. It is less-known that a linearly polarized excitation creates in graphene-like materials a strongly anisotropic distribution of photoexcited carriers, with their momenta aligned preferentially normal to the polarization plane, allowing one to effectively steer the direction of electrons by light~\cite{BookHartmann2011,*thesishartmann}. This largely overlooked effect is the central theme of our paper. A similar momentum alignment phenomenon occurs in bulk GaAs-type semiconductors~\cite{Zakharchenya_84} and quantum wells \cite{Merkulov1991,Mirlin_SST_92,Kash92PRL,PortnoiSovPhysSemicond91,PortnoiSovPhysSemicond92}, where the electrons created by the interband absorption of linearly polarized light are also distributed anisotropically in momentum space; the same selection rules govern polarization properties  of quantum-well-based lasers \cite{ShtengelSemiSci93}. In conventional semiconductors the alignment is due to the spin-orbit interaction, whereas in graphene, it is due to the pseudo-spin. Namely, the ratio of the two components of the spinor-like graphene wavefunction depends on the direction of momentum which influences the optical transition selection rules. Unlike semiconductors where most optical phenomena are associated with the band-edge transitions, photoexcited carriers in graphene are always created with a significant value of momentum and a linearly polarized excitation results in strong momentum anisotropy for all excitation energies.
Momentum alignment in graphene was first discussed in Ref.~\cite{BookHartmann2011,*thesishartmann} followed by a number of papers mostly dealing with the photogalvanic effect see, e.g., Ref.~\cite{Durnev2021} and references therein. Notably, in this paper we do not deal with photocurrents, which are largely influenced by relaxation processes~\cite{Golub11}, rather, we focus on the selection rules and the shape of the momentum distribution function at the instant of photoexcitation.

In graphene and graphene-like two-dimensional (2D) Dirac materials, such as single layers of group-VI dichalcogenides \cite{PhysRevLett.108.196802,Falko_material_15} and elemental analogues of graphene \cite{Madhu_Small_15,PhysRevB.96.075427} including silicene, germanene, stanene and several others, the electronic properties can be described in terms of particles belonging to two valleys \cite{schaibley2016valleytronics}, centered around the symmetry points \textbf{K} and $\textbf{K}^{\prime}$. These points are nonequivalent and degenerate in terms of energy. This degree of degeneracy is the so-called valley degree of freedom. It has been proposed that this additional quantum number can be utilized in an analogous manner to spin in semiconductor spintronics \cite{Zutic_04} and has been suggested as a basis for carrying information in graphene-based devices \cite{Rycerz_Nature_07}. 
For gapped 2D Dirac materials, excitation by linearly polarized light with a photon energy just above the band gap results in an equal population of both valleys.
However, as we show below, at higher excitation energies, the trigonal warping effect (an anisotropy of the equal-energy contour) in conjunction with momentum alignment can be utilized to spatially separate carriers belonging to different valleys, thus providing a route to optovalleytronics - the optical control of valley population. This optical valley Hall effect becomes stronger as the photon energy increases.

In gapless 2D materials, such as graphene, in addition to the aforementioned trigonal warping which occurs at high energies, there is a strong anisotropic modification to the spectra near the apex of the Dirac cone due to the celebrated Rashba spin-orbit effect~\cite{Rashba1960,Rashba1984,Bihlmayer2015}, which is inevitable in the presence of a substrate and can be controlled by the back-gate voltage.
The importance of Rashba spin-orbit interaction for graphene physics was realized by Kane and Mele~\cite{Kane2005} practically simultaneously with graphene exfoliation which was followed by an extensive body of research including important contributions by Rashba himself~\cite{Rashba2009,Rashba2009b,Marchenko2012}.
The presence of the Rashba term, which is significant at low energies, should result in an optical valley Hall effect occurring also at much lower far-infrared frequencies. 


In contrast to the optical spin and valley Hall effects in polaritonics~\cite{PhysRevLett.95.136601,leyder2007observation,PhysRevB.96.165432}, our predicted optical valley Hall effect, caused by the spatial separation of carriers belonging to different valleys by linearly polarized light, does not need a microcavity. A better-known alternative route to optovalleytronics utilizes circularly polarized light in quasi-2D Dirac materials with non-zero effective mass. Unlike in gapless 2D Dirac materials, the selection rules for interband transitions for circularly polarized light in gapped graphene-like systems are strongly valley dependent. Namely, circularly polarized radiation of a specific handedness, with an energy matching the band gap, will excite electrons in one valley only. These optical transition selection rules are independent of the physical nature of the gap. The gap can be opened, e.g., by placing graphene on a matching substrate with two chemically different atoms underneath the two neighbouring carbon atoms~\cite{Giovannetti2007,Zhou2007} or chemically functionalized graphene~\cite{Craciun2013}. The valley-dependent selection rules for circularly polarized light can be utilized for the detection of the optical valley Hall effect.
In single layers of transition metal dichalcogenides~\cite{Manzeli2017} the gap is believed to be of a mixed nature, involving significantly differing on-site energies accompanied by spin-valley locking due to strong spin-orbit coupling. Thus, in these materials the optical valley Hall effect is automatically accompanied by the optical spin Hall effect. 


The momentum alignment phenomenon also has profound consequences for optical transition selection rules in narrow-gap carbon nanotubes (CNTs) and graphene nanoribbons (GNRs). Namely, the angular dependence of graphene's momentum distribution function leads to a spectacular enhancement in the matrix element of optical interband transition at the band edge in these quasi-one-dimensional (1D) nanostructures. This enhancement is due to an effective quantized momentum in the direction normal to the CNT or GNR axis which governs the interband transitions when the free momentum along the axis is small. The pronounced peak in the optical matrix element at the band gap edge has a universal value which is proportional to the Fermi velocity and is independent of nature of the gap, which can be magnetic-field, curvature or edge-effect induced.

In what follows, we derive the optical selection rules for interband transitions in 2D Dirac materials for both linearly and circularly polarized excitations. In the low-energy regime and in the absence of the Rashba term, the optical selection rules for linearly polarized excitations are shown to be valley independent; the same is true for circularly polarized excitations in graphene. In contrast, for gapped 2D Dirac materials the optical transitions associated with circularly polarized light are valley-dependent. The distribution of photoexcited carriers is first calculated in the absence of the Rashba term, for both the low-energy regime and for the range of frequencies in which trigonal warping effects become important. Next, we show that in the presence of warping, a linearly polarized excitation will result in the spatial separation of carriers belonging to different valleys (optical valley Hall effect). An experimental set up is proposed to observe this effect in gapped 2D Dirac materials. 

We then return to the low-energy part of the spectrum of gapless materials, but with the Rashba term introduced. This term results in a strong warping effect near the apex of the Dirac cone. This leads to a spectacular valley separation effect upon a linearly polarized excitation, at a completely different photon energy scale. This energy scale can be tuned by modifying the strength of the Rashba term by changing the value of the back-gate voltage.

We also apply the theory of momentum alignment in graphene to quasi-1D systems such as narrow-gap CNTs and armchair GNRs (AGNRs). These systems are shown to have strong low-energy interband transitions, which are typically in the THz range~\cite{Portnoi2015}.  Finally, we discuss the possibility for the experimental observation of the predicted strong THz transitions and how they could be used in THz emitters.

\section{Optical selection rules for 2D Dirac materials}\label{sec:Rules}
In the dipole approximation the transition rate of an electron, of
wave vector $\boldsymbol{k}$, from the conduction to the valence band
is given by Fermi's Golden Rule:
\begin{equation}
W_{\boldsymbol{k}}=\frac{2\pi e^{2}I_{\boldsymbol{e}}}{ch\nu^{2}}
\left|\boldsymbol{e}\cdot\left\langle \Psi_{1}\left(\boldsymbol{k}\right)\left|
\hat{\boldsymbol{v}}
\right|\Psi_{2}\left(\boldsymbol{k}\right)\right\rangle \right|^{2}
\delta\left(\xi_{1}-\xi_{2}-h\nu\right),
\label{eq:Transition}
\end{equation}
where $\Psi_{j}$ ($j = 1,2$) are the eigenfunctions of the electrons
in the conduction and valence bands and $\xi_{1}$, $\xi_{2}$ are their
associated energies.  Here $\hat{\boldsymbol{v}}$, $I_{\mathbf{e}}$
and $\nu$ are the velocity operator, intensity
and frequency of the excitation respectively; $\boldsymbol{e}$ is the polarization of
the excitation, which we assume to propagate normal to the crystal's surface. The tight-binding Hamiltonian of a gapped graphene-like 2D crystal can be written as:
\begin{equation}
\label{eq:HamiltonianWithSOCGeneral}
H \left( \boldsymbol{k} \right) = \left(\begin{array}{cc}
\frac{1}{2}E_{g}
& f\left(\boldsymbol{k}\right)\\
f^{\ast}\left(\boldsymbol{k}\right)
& -\frac{1}{2}E_{g}
\end{array}\right) \, ,
\end{equation}
where $E_{g}$ is the band gap; $f\left(\boldsymbol{k}\right)=\sum_i t_i \exp\left(i \boldsymbol{k} \cdot \boldsymbol{R}_i\right)$ where $\boldsymbol{R}_i$ are the nearest neighbor vectors and $t_i$ are their associated hopping integrals~\cite{saito1998physical}. This Hamiltonian acts on the basis $\left(\left|\psi_{A}\right\rangle ,\,\left|\psi_{B}\right\rangle \right)^{T}$, where $\left|\psi_{A}\right\rangle$ and $\left|\psi_{B}\right\rangle$ are the tight-binding wavefunctions associated with the two sub-lattices. In pristine graphene, the band gap due spin-orbit coupling has been measured to be $42.2$~$\mu$eV~\cite{Sichau2019}, and as such  $E_{g}$ is negligibly small. However, the gap can be opened in a variety of ways; indeed this Hamiltonian can be used to describe graphene on a staggered sublattice potential~\cite{Yao2008} such as boron nitride~\cite{Giovannetti2007} or SiC~\cite{Zhou2007,Mattausch2007}, where the gap is three orders of magnitude larger. The Hamiltonian, Eq.~(\ref{eq:HamiltonianWithSOCGeneral}), admits the eigenvalues
\begin{equation}
\xi_{j} = (-1)^{j+1} \sqrt{\frac{1}{4}E_{g}^2 + |f|^2} \, ,
\label{eq:EigenvaluesSO}
\end{equation}
where $j=1,2$ correspond to the conduction band and valence band, respectively.
Their corresponding eigenvectors are given by
\begin{equation}
\label{eq:EigenvectorsSO}
\boldsymbol{\chi}_{j} =
C_{j}
\left(\begin{array}{c}
1\\
\frac{\xi_{j}-\frac{1}{2}E_{g}}{f}
\end{array}\right)
\, ,
\end{equation}
where $|C_{j}|^2=(\xi_j+\frac{1}{2}E_{g})/(2\xi_j)$. The eigenfunctions entering Eq.~(\ref{eq:Transition}) are given by $\Psi_{j}=\chi_{A, j}\left|\psi_{A}\right\rangle + \chi_{B,j}\,\left|\psi_{B}\right\rangle$. Within the gradient approximation, the velocity operator matrix can be written as
\begin{equation}
\label{eq:VelocityOperatorSO}
\hat{\boldsymbol{v}} = \frac{1}{\hbar}\frac{\partial H}{\partial \boldsymbol{k}} =
\frac{1}{\hbar}
\left( \begin{array}{cc}
	0 &  \nabla_{\boldsymbol{k}}f\\
	 \nabla_{\boldsymbol{k}}f^{\ast} & 0
\end{array}\right) \, .
\end{equation}
Using Eqs.~\eqref{eq:EigenvaluesSO}, \eqref{eq:EigenvectorsSO} and~\eqref{eq:VelocityOperatorSO} allows the matrix element of velocity to be expressed as
\begin{equation}
\langle\Psi_{1}\left|\hat{\boldsymbol{v}}\right|\Psi_{2}\rangle=-\frac{1}{\hbar|f|}\left[\frac{E_{g}}{\sqrt{E_{g}^{2}+4|f|^{2}}}\Re\left(f^{\ast}\nabla_{\boldsymbol{k}}f\right)+i\Im\left(f^{\ast}\nabla_{\boldsymbol{k}}f\right)\right]\,.
\label{eq:VMEso}
\end{equation}
As can be seen from Eq.~\eqref{eq:VMEso}, the band gap affects only the real part of the matrix element, which, due to the peculiar form of the factor $\tfrac{E_{g}}{\sqrt{E_{g}^{2}+4|f|^{2}}}$, has a sharp dependence on the electron momentum near the Dirac point where $f \rightarrow 0$.

\subsection{The low-energy regime}
We will now consider the low-energy regime where the effective matrix Hamiltonian,
Eq.~(\ref{eq:HamiltonianWithSOCGeneral}), can be approximated by a first order Taylor series
expansion in $\boldsymbol{\kappa}$, which is defined as the momentum measured relative
to the Dirac point of interest, i.e. the ${\bf K}$ or ${\bf K}^{\prime}$ point. The results of this subsection can be obtained by the formal low-energy expansion of Eq.~(\ref{eq:VMEso}). However, due to the importance of the conical approximation for graphene and the parabolic limit for gapped Dirac systems, it is instructive to consider the low-energy regime starting from the well-known approximate matrix Hamiltonian. For definiteness we will use the two nonequivalent Dirac points as ${\bf K}=\left(0,-4\pi/3a\right)$ and ${\bf K}^{\prime}=-{\bf K}$, where $a$ is the lattice constant, and we employ the same coordinate system as in Fig. 4.1 of Ref. \cite{saito1998physical}. The low-energy expansion of the Hamiltonian given by Eq.~\eqref{eq:HamiltonianWithSOCGeneral} in the vicinity of the two nonequivalent Dirac points can be written as the effective Hamiltonian \begin{equation}
\label{eq:HamiltonianWithSOC}
H \left( \boldsymbol{\kappa} \right) =
\hbar v_F \left(\begin{array}{cc}
\Delta & \kappa_x + i (-1)^s \kappa_y\\
\kappa_x - i (-1)^s \kappa_y & -\Delta
\end{array}\right) \, ,
\end{equation}
which acts on the basis $\left(\left|\psi_{A}\right\rangle ,\,-i\left|\psi_{B}\right\rangle \right)^{T}$. Here $\Delta = E_{g} / (2\hbar v_F)$, $\kappa_{x,y}$ are the Cartesian components of $\boldsymbol{\kappa}$, and $s$ is the valley index number, which has a value of 1 or 2 for the {\bf K} or {\bf K}$^{\prime}$, respectively. This Hamiltonian can also be used to model the low-energy band structure of low-buckled silicene, germanene and tinene~\cite{Liu2011}. In these systems the spin-orbit interaction opens up a narrow-band gap of the order of several meV~\cite{Liu2011}. The Hamiltonian~\eqref{eq:HamiltonianWithSOC} has the following eigenvalues:
\begin{equation}
\varepsilon_{j} = (-1)^{j+1} \hbar v_\mathrm{F} \sqrt{\Delta^2 + \kappa^2} \, ,
\label{eq:EigenvaluesSOCLowEnergy}
\end{equation}
where $\kappa^2 = \kappa^2_x + \kappa^2_y$. The corresponding normalized conduction and valence band wave functions are
\begin{equation}
\label{eq:EigenvectorsWithSOC1}
\chi_{j}=
D_j
\left(\begin{array}{c}
1\\
\sqrt{\frac{\varepsilon_{j}-\frac{1}{2}E_{g}}{\varepsilon_{j}+\frac{1}{2}E_{g}}}
e^{-i\left(-1\right)^{s}\varphi_{k}}
\end{array}\right)
\end{equation}
where $|D_{j}|^2=(\varepsilon_j+\frac{1}{2}E_{g})/(2\varepsilon_j)$ and $\tan\varphi_{\kappa}=\kappa_{y}/\kappa_{x}$.

Let us now consider the perturbation introduced into the Hamiltonian by an incident electromagnetic wave. The wave can be described by the vector potential:
$\boldsymbol{A}=cE\left( a_x \boldsymbol{e}_x + a_y \boldsymbol{e}_y \right) e^{-i \omega t}/(2\omega)$
where $c$ is the speed of light, $\omega$ is the wave's frequency and $E$ is the magnitude of the electric field strength. The parameters $a_x$ and $a_y$ can be defined as $a_x = \cos \phi $ and $a_y= \sin \phi$ for linearly polarized light and $a_x = 1$ and $a_y = \mp i$ for circularly polarized light. 
The angle $\phi$ denotes the angle between the polarization vector and the $x$-axis. The circularly polarized wave is considered as the sum of two waves polarized along the $x$ and $y$ axis with a phase difference of $\pi/2$ between the components. 
The `$+$' and `$-$' signs correspond to left- and right-handed circular polarizations respectively. In the presence of an electromagnetic field, the particle's modified momentum is found by using minimal substitution, $\boldsymbol{\kappa} \rightarrow \boldsymbol{\kappa} + e \boldsymbol{A} / \hbar c$, from which the gradient approximation used in Eq.~\eqref{eq:VelocityOperatorSO} stems (see discussion after Eq.~(36) in Ref.~\cite{Saroka2017}). Applying this minimal substitution to the Hamiltonian~\eqref{eq:HamiltonianWithSOC} allows for linearly polarized light to express the dot product of the polarization vector $\boldsymbol{e}$ and $\hat{\boldsymbol{v}}$ as:
\begin{equation}
\label{eq:LinearPolarizationWithSOC}
\boldsymbol{e}\cdot \hat{\boldsymbol{v}}=
v_F \left(\begin{array}{cc}
0 & e^{i(-1)^s \phi}\\
e^{-i(-1)^s \phi} & 0
\end{array}\right) \,.
\end{equation}
For circularly polarized light the relevant matrix is
\begin{equation}
\label{eq:CircularPolarizationWithSOC}
\hat{v}_x + i(-1)^{p}\hat{v}_y=
v_F \left(\begin{array}{cc}
0 & 1 - (-1)^{p+s}\\
1+(-1)^{p+s} & 0
\end{array}\right) \, ,
\end{equation}
where $p=1$ and $p=2$ for right and left-handed circularly polarized light respectively. Notably, the matrix given by Eq.~(\ref{eq:CircularPolarizationWithSOC}) has only one non-zero element which leads to important consequences for the optical selection rules in gapped materials.

The angular generation density $g\left(\varphi_{\kappa}\right)$, which is the angular dependence of the momentum distribution function of photoexcited carriers in the absence of any relaxation, is defined such that
$g\left(\varphi_{\kappa}\right)d\varphi_{\kappa}$ gives the
number of carriers per unit area created in the angle range $\varphi_{\kappa}$
to $\varphi_{\kappa}+d\varphi_{\kappa}$.
For a single valley and spin, the angular generation density is given by
\begin{eqnarray}
\label{eq:angulargen}
  g\left(\varphi_{\kappa}\right) &=&
  \left(\frac{1}{2\pi}\right)^{2}\intop
W_{\boldsymbol{k}}\left(\kappa,\varphi_{\kappa}\right)\kappa d\kappa.
\end{eqnarray}
For linearly polarized light,
\begin{equation}
\label{eq:DistributionFunctionOfPhotoExcitedCarriers}
\left|\boldsymbol{e}\cdot\left\langle \Psi_{1}\left|
\hat{\boldsymbol{v}}
\right|\Psi_{2}\right\rangle \right|^{2}=
v_{\mathrm{F}}^2 \left[ \sin^2 \theta + \dfrac{\Delta^2}{\kappa^2 + \Delta^2} \cos^2 \theta \right],
\end{equation}
where $\theta = \varphi_{\kappa} -\phi$ is the angle between the momentum of the photoexcited electron and the polarization of the incident light. Therefore, the angular generation density of a gapped graphene-like crystal is
\begin{equation}
g\left(\varphi_{\kappa}\right)=\frac{e^{2}} {4\hbar
c}\frac{I_{e}}{h\nu}
\left[\sin^{2}
\left(\varphi_{\kappa} -\phi\right)
+\dfrac{E_{g}^2}{h^{2}\nu^{2}}\cos^{2}
\left(\varphi_{\kappa} -\phi\right)
\right].
\label{eq:trans_prob}
\end{equation}
Since Eq.~(\ref{eq:DistributionFunctionOfPhotoExcitedCarriers}) is valley-independent, the  angular generation density is the same for both valleys. It also follows from Eq.~\eqref{eq:DistributionFunctionOfPhotoExcitedCarriers} that for non-zero $\Delta$, when $\kappa \ll \Delta$ there is no momentum alignment of the photoexcited carriers in gapped 2D Dirac systems and the optical matrix element has a universal value, independent of the gap, and proportional to the Fermi velocity. By integrating Eq.~(\ref{eq:trans_prob}) over all directions of electron momentum and summing over valley and spin indices, one can easily recover the results obtained within the optical conductivity formalism for graphene~\cite{Kuzmenko_PRL_08,Koshino_PRB_08,Ryzhii_JAP_07,Stauber_PRB_08} including the universal value of single layer opacity~\cite{Nair_Science_08}.

\begin{figure}
    \centering
    \includegraphics[width=15cm]{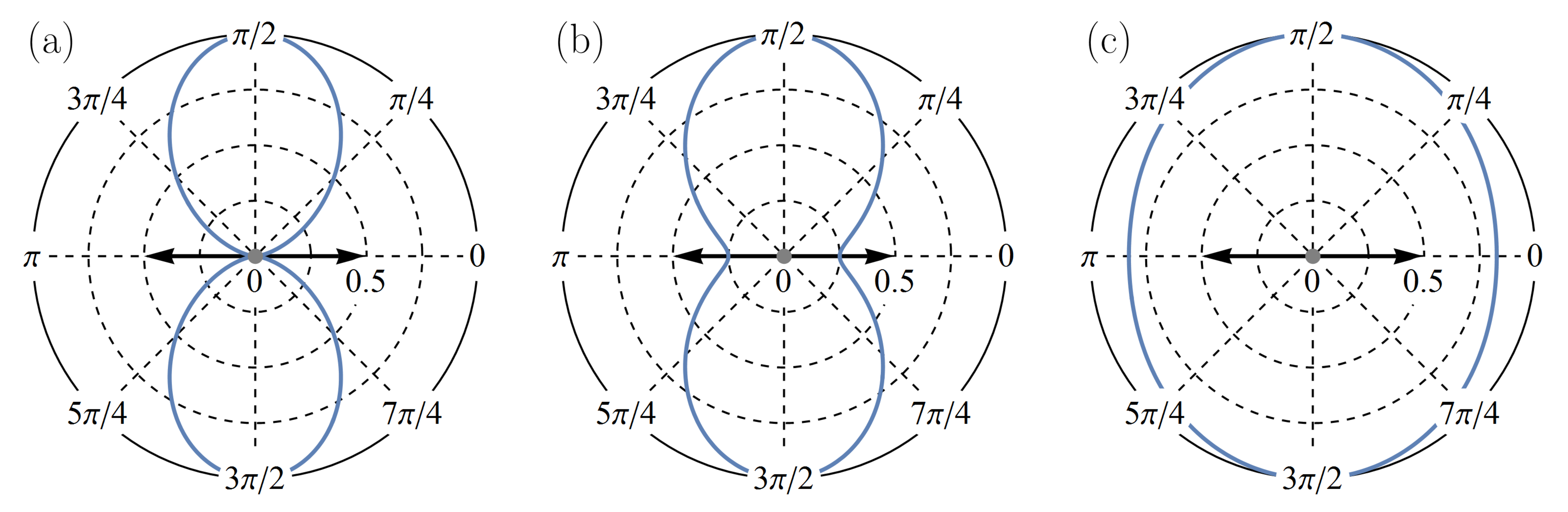}
\caption{The polar plots of the momentum distribution functions for (a) graphene and an isotropic gapped 2D Dirac material excited at (b) $h \nu = 2 E_{g}$ and (c) $h \nu = 1.1 E_{g}$, for $\phi=0$. Here, the black arrow represents the polarization of the excitation, which is assumed to be along the x-axis.
}
\label{fig:LER}
\end{figure}

It is convenient to rewrite Eq.~(\ref{eq:trans_prob}) as
\begin{equation}
g\left(\varphi_{\kappa}\right)=F_{0}\left(\varepsilon_{0}\right)\left[1+\alpha_{0}\cos\left(2\theta\right)\right]
\, ,
\label{eq:Distribution0}
\end{equation}
where $\alpha_{0}=\left(E_{g}^2-4\varepsilon^{2}_0\right)/\left(4\varepsilon^{2}_0+E_{g}^2\right)$ defines the degree of momentum alignment, $F_0(\varepsilon_0)$ gives the total density of carriers created at energy $\varepsilon_0$, and the subscript 0 reflects the fact that no relaxation has occurred. It should be noted that after relaxation the momentum distribution function in 2D retains the same angular dependence~\cite{portnoi1993depolarization,Kainth2002}. Unlike conventional semiconductor quantum wells~\cite{merkulov1991theory,Merkulov1991}, in graphene's low-energy regime $\alpha_{0}=-1$ and is not a function of the excitation energy.  The illumination of graphene with linearly polarized radiation results in an anisotropic momentum distribution of photoexcited carriers, with the preferential direction of propagation normal to the excitation's polarization plane. In Fig.\,\ref{fig:LER}, polar plots of $g\left(\varphi_{\kappa}\right)$ are shown for light of normal incidence and the polarization angle $\phi=0$ for graphene (Fig.\,\ref{fig:LER}(a)) and for an isotropic gapped 2D material excited at $h \nu = 2 E_{g}$ (Fig.\,\ref{fig:LER}(b)) and $h \nu = 1.1 E_{g}$ (Fig.\,\ref{fig:LER}(c)).

\begin{figure}
  \centering
    \includegraphics[width=10cm]{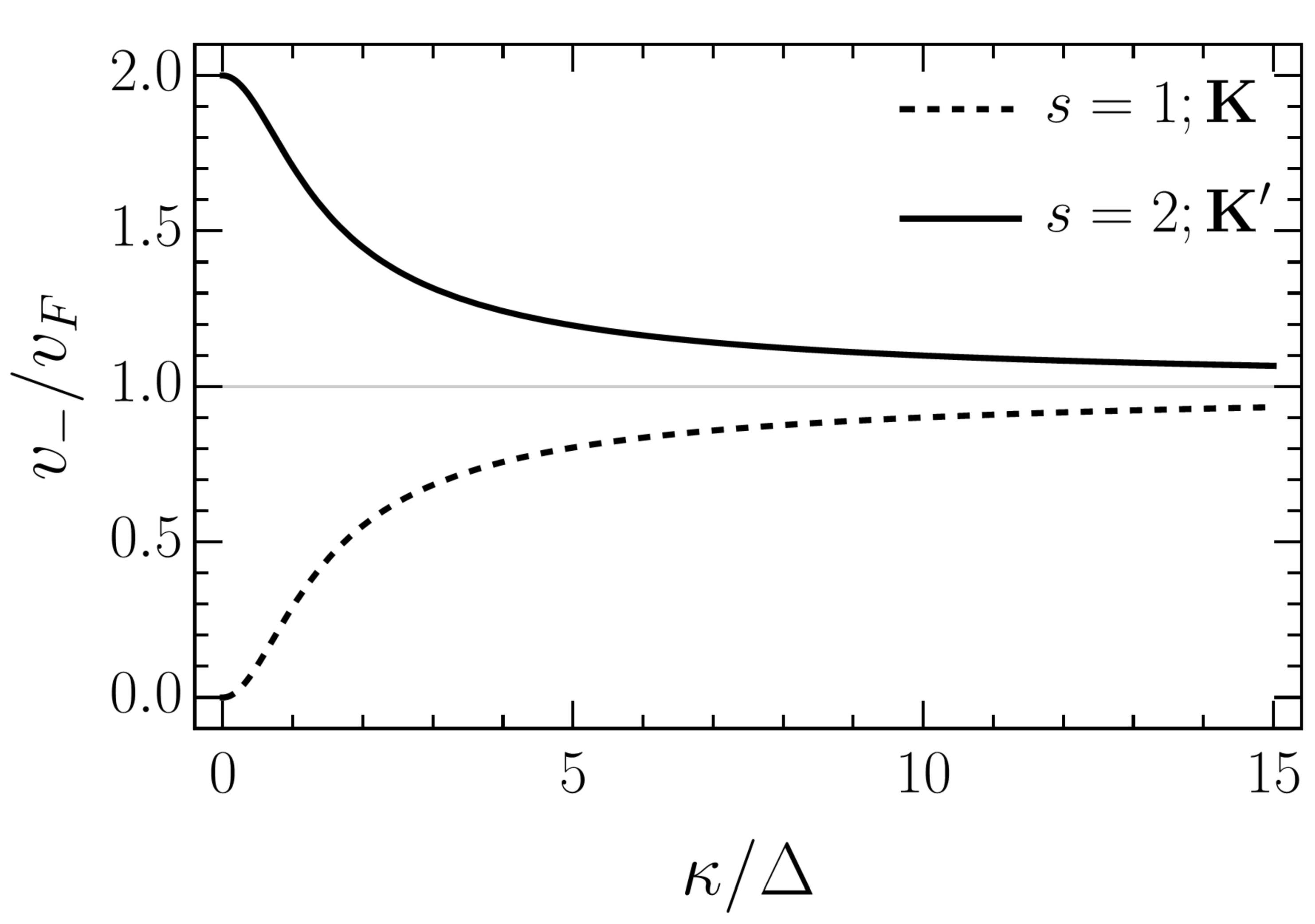}
    \caption{
The absolute value of the matrix element of the velocity operator associated with interband transitions induced by right-handed circularly polarized light as a function of excitation energy. The lower and upper curves correspond to the \textbf{K} and $\textbf{K}^{\prime}$ valleys, respectively.
    }
\label{fig:cir_matrix}
\end{figure}

The  optical  selection  rules  for  interband transitions resulting from linearly  polarized  light,  of  low  excitation energies in 2D  Dirac  materials  are  valley  independent. However, in what follows we shall see that the optical selection rules of interband transitions caused by circularly polarized light in gapped graphene-like crystals are strongly valley dependent. Evaluating the matrix elements for the velocity operator projections given by Eq.~\eqref{eq:CircularPolarizationWithSOC} yields for right- and left-handed circularly polarized light
\begin{align}
\label{eq:CircularPolarizationRKvalley}
v_{-}=
\left|
\langle\Psi_{1}\left|\hat{v}_x-i\hat{v}_y\right|\Psi_{2}\rangle
\right|
=
v_{F} \left|
\left(-1\right)^{s}+\dfrac{\Delta}{\sqrt{\kappa^{2}+\Delta^{2}}}
\right| \, ,
\end{align}
and
\begin{align}
\label{eq:CircularPolarizationLKvalley}
v_{+}=
\left|
\langle\Psi_{1}\left|\hat{v}_x+i\hat{v}_y\right|\Psi_{2}\rangle
\right|= v_{F} \left|
\left(-1\right)^{s}-\dfrac{\Delta}{\sqrt{\kappa^{2}+\Delta^{2}}}
\right| \, .
\end{align}
In striking contrast to graphene, where both circular polarizations lead to the equal occupation of both valleys, the valley degree of freedom in gapped 2D Dirac materials can be manipulated by the polarization of the circularly polarized light. For $s=1$ and as $\kappa \rightarrow 0$ the matrix element of velocity given by Eq.~\eqref{eq:CircularPolarizationRKvalley} vanishes while in Eq.~\eqref{eq:CircularPolarizationLKvalley} it does not.
Thus, when the excitation energy matches the band gap, only the left-handed polarization can excite interband transitions for the {\bf K} valley: $v_+=2v_{F}$ and $v_-=0$.  For large excitation energies, the matrix element of velocity $v_+$ decreases with increasing frequency until it attains a half of its band-edge value, whereas $v_-$ for the same valley increases towards $v_{F}$.
By contrast, for the {\bf K}$^{\prime}$ valley, i.e. for $s=2$, only the right-handed polarization leads to band-edge transitions.
Therefore, by exciting gapped graphene-like materials with circularly polarized light of near-band-edge photon energy one can achieve full valley polarization.
The valley selection rules for band-edge transitions are indeed well-known mostly due to the proliferation of research in the optics of single-layer transition metal dichalcogenides (TMDs)~\cite{PhysRevLett.105.136805,xu2014spin}. However, it seems to be overlooked in the analysis of TMDs experiments that the valley polarization by circularly polarized light diminishes when the excitation photon energy significantly exceeds $E_g$. This reduction in valley polarization can be clearly seen from Fig~\ref{fig:cir_matrix}, in which we show in the same plot $v_{-}/v_{F}$ for two nonequivalent valleys. The population of both valleys becomes the same for $h\nu \gg E_g$; in this limit the optical properties of our idealized gapped material would mimic those of graphene including its universal absorption.

\subsection{The trigonal warping regime}
\begin{figure}
  \centering
    \includegraphics[width=12cm]{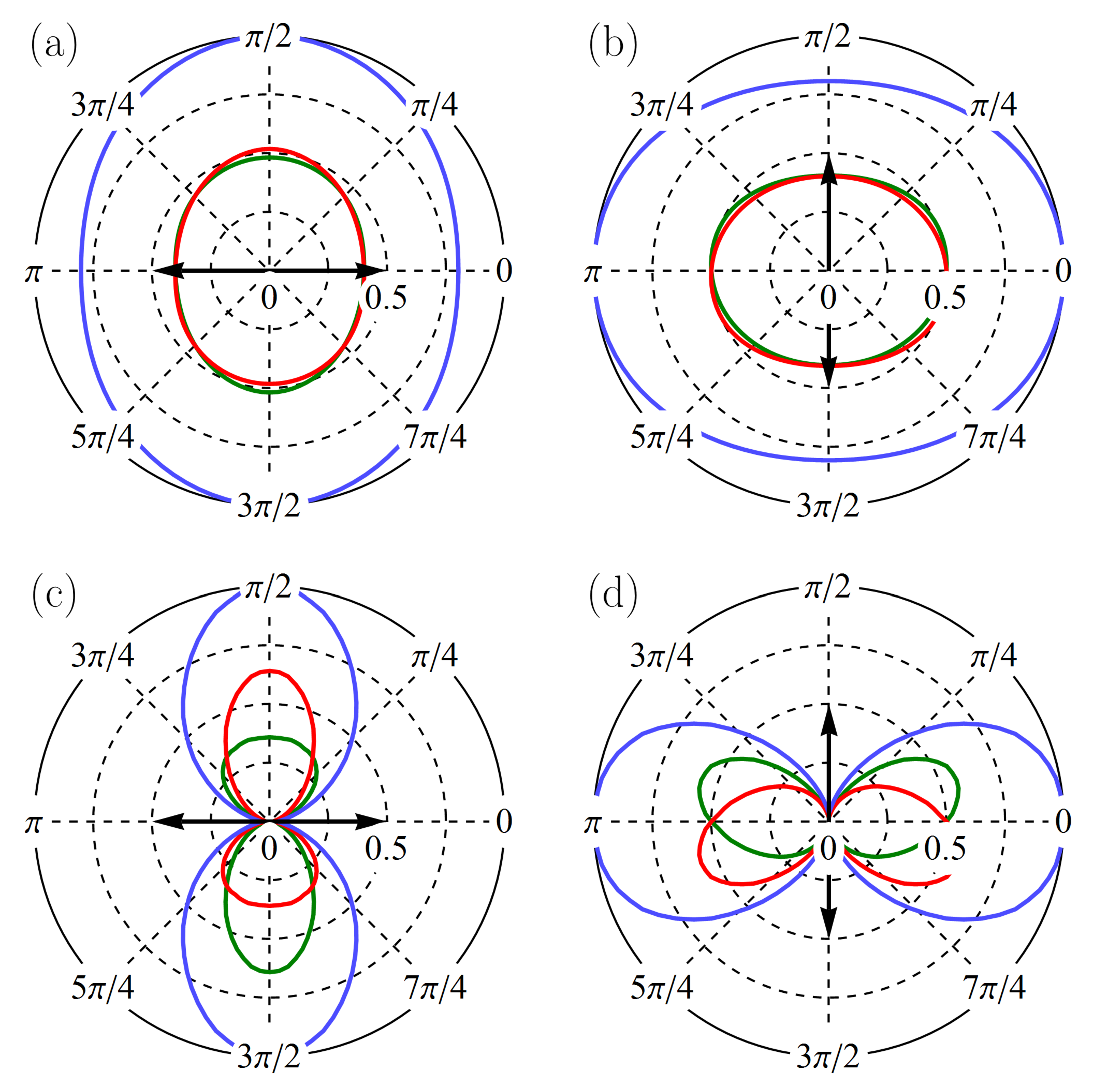}
    \caption{The polar plots of the momentum distribution function of photoexcited carriers generated by linearly polarized light in a gapped graphene-like crystal when the trigonal warping is taken into account for: (a) $\phi=0$, $h\nu= 1.1\ E_g$ and $E_g = 0.1\, |t|$, (b) $\phi=\pi/2$, $h\nu= 1.1\, E_g$ and $E_g = 0.1\, |t|$, (c) $\phi=0$, $h\nu= 5\, E_g$ and $E_g = 0.1\, |t|$, (d) $\phi=\pi/2$, $h\nu= 5\, E_g$ and $E_g = 0.1\, |t|$.  Here $\phi$ is the angle between the excitation polarization plane and the $x$-axis at normal light incidence.
    The energy gap is given in terms of the hopping integral, $|t| \approx 3$~eV. The green and red lines show the contributions from the  {\bf K} and {\bf K}$^{\prime}$ valleys, respectively, while the blue contours are their sum, and the bold black arrows represent the polarization of the excitation.
    }
\label{fig:HER}
\end{figure}

Our analysis of optical transitions induced by polarized light is also applicable everywhere away from the Dirac cone. However, the most interesting regime is when the isotropic approximation becomes no longer valid, but the equienergy contours for photoexcited carriers remain as closed curves (shaped as triangles with smooth corners) around the two $\textbf{K}$ points. At these energies, i.e. in the trigonal warping regime, the dispersion is no longer linear, rather it becomes $2\pi/3$ periodic in $\phi_{\kappa}$. This new symmetry allows for the optical control of valley polarization by linearly polarized light. Combining Eqs.~(\ref{eq:Transition}) and (\ref{eq:angulargen}) yields for the angular generation density in a given valley:
\begin{equation}
g_{s}\left(\varphi_{\kappa}\right)=\frac{e^{2}I_{\boldsymbol{e}}}{4\pi^{2}c\hbar\nu^{2}}\intop\left|\boldsymbol{e}\cdot\left\langle
\Psi_{1}
\left|\hat{\boldsymbol{v}}\right|
\Psi_{2}
\right\rangle \right|^{2}\delta\left(2\xi_{1}\left(\kappa,\varphi_{\kappa}\right)-h\nu\right)
\kappa d\kappa ,
\label{eq:distributionfunc}
\end{equation}
where the matrix element, $\left\langle \Psi_{1}\left|\hat{\boldsymbol{v}}\right|\Psi_{2}\right\rangle$, entering Eq.~(\ref{eq:distributionfunc}) is given by Eq.~(\ref{eq:VMEso}).
For simplicity we removed the valley index $s$ from the right hand side of Eq.~(\ref{eq:distributionfunc});
however, $\xi_1\left(\kappa,\varphi_{\kappa}\right)$ and $f$ are indeed valley-dependent.
In the trigonal warping regime the contributions to the momentum distribution function from the {\bf K} and {\bf K}$^{\prime}$ valleys are not equivalent. The contribution from an individual valley does not preserve the necessary symmetry of the momentum distribution function of photoexcited charge carriers which should reflect the absence of net photocurrent at normal light incidence in a uniform system without an applied static external potential. Only when both valleys are taken into account the symmetry of $g\left(\varphi_{\mathbf{\kappa}}\right)=g_1\left(\varphi_{\mathbf{\kappa}}\right)+g_2\left(\varphi_{\mathbf{\kappa}}\right)=g\left(\varphi_{\mathbf{\kappa}}+\pi\right)$ is restored, as can be seen from Fig.~\ref{fig:HER}. Therefore two \textbf{K} points are required to describe the optical properties of the system and that as soon as trigonal warping becomes important the one-valley picture provides an insufficient description.
The majority of publications on graphene are restricted to a single-valley picture, although inter-valley scattering and valley mixing are essential for explaining weak localization data \cite{McCann_PRL_07,Tikhonenko_PRL_08}
and are used in the study of GNRs
\cite{Brey_PRB_06}. The effect discussed here is an example of a phenomenon in a pristine infinite graphene-like system where disregarding one of the two valleys gives an unphysical result.
\begin{figure}
    \centering
    \includegraphics[width=10cm]{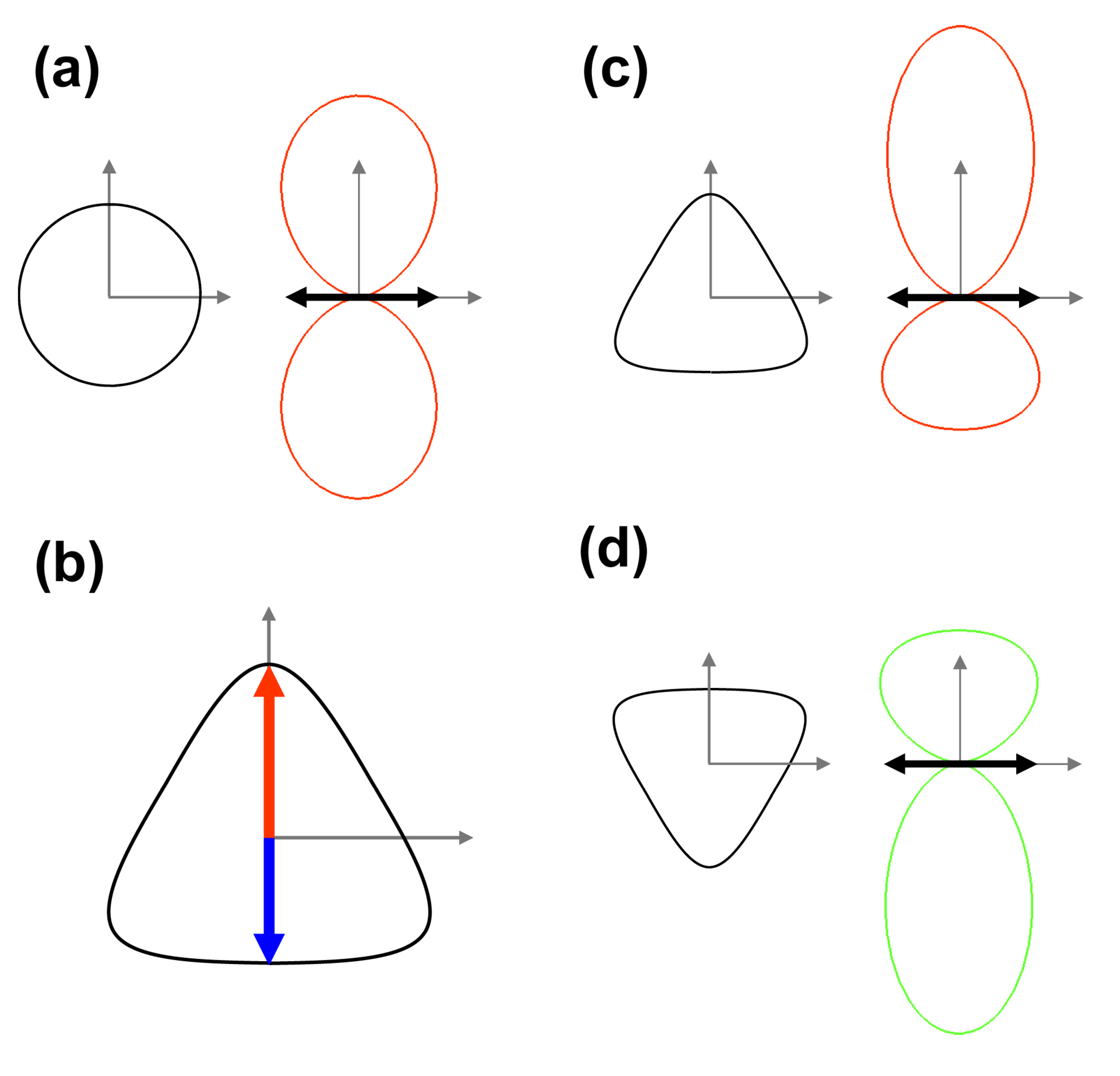}
    \caption{
(a) The polar plots of the momentum distribution of photoexcited carriers in the low-energy regime at normal light incidence when the excitation polarization plane is along the $x$-axis, i.e. $\phi=0$. (b) The equienergy contour around the $\textbf{K}^{\prime}$ point (black line), in the regime where trigonal warping becomes important. Here, the red line indicates a direction in k-space where $\kappa$ is maximum, while the blue line indicates a direction in k-space where $\kappa$ is minimum. The contribution to the momentum distribution of photoexcited carriers from the (c) $\textbf{K}^{\prime}$ point and (d) \textbf{K} point is shown. In both instances $\phi=0$ and $h\nu = 0.6 |t|$. As a guide to the eye, the corresponding equienergy contours are drawn next to each distribution function and the bold black arrows represent the polarization of the excitation. 
}
\label{fig:ang_DOS}
\end{figure}

The symmetry breaking between the {\bf K} and {\bf K}$^{\prime}$ valley can be understood as follows. The conservation of energy during an optical transition maps out an equienergy contour in $k$-space which represents all the allowed values of $\boldsymbol{\kappa}$ taking part in the transition. In the low-energy regime this contour is a circle. However, in the trigonal warping regime, the magnitude of the allowed $\boldsymbol{\kappa}$-vectors is angle-dependent, and the equienergy contours are triangular in shape (see Fig.~\ref{fig:ang_DOS}). The length of vector $\boldsymbol{\kappa}$ reaches its maximum at the angles directed towards the apices of the triangle and is shortest in the directions towards the triangle sides. Since the angular generation density is dependent on the size of the integral element, $\kappa d\kappa$, the density of the excited carriers is maximized along those angles directed towards the apices of the triangles. This is manifested in the momentum distribution by the elongation of the lobes along such directions. Thus, each valley carries excited particles with different preferential angles of momentum. The valley symmetry breaking seen in Fig.~\ref{fig:HER} occurs because the apices of the equienergy contours are in opposite directions for the two nonequivalent valleys. This eventually leads to the imbalance in the number of carriers belonging to two different valleys traveling in different directions, e.g. up and down in Fig.~\ref{fig:HER}.
\begin{figure}
  \centering
    \includegraphics[width=0.9\linewidth]{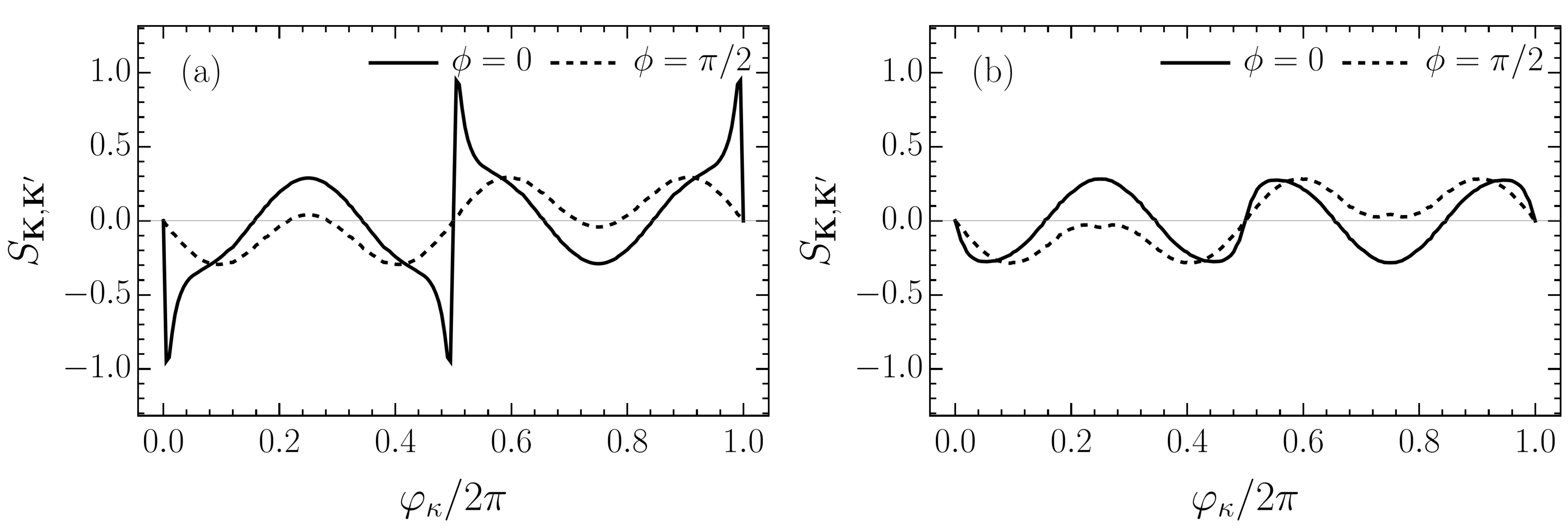}
\caption{
The degree of valley polarization for (a) $E_g = 0$ and (b) $E_g = 0.1\, |t|$ for an excitation frequency of $h\nu= 0.5\, |t|$. The solid and dashed lines correspond to the polarization angles $\phi=0$ and $\phi=\pi/2$ respectively.
}
\label{fig:Valley_Polar}
\end{figure}
Within the trigonal warping regime a linearly polarized excitation could be utilized as a control mechanism in valleytronic devices. It can be seen from Fig.~\ref{fig:HER} that for a given polarization, the asymmetry between each valley's contribution to the momentum distribution is maximized for preferential directions. Along such directions there is a net difference between carriers generated from the {\bf K} and {\bf K}$^{\prime}$ valley. To quantify this effect we introduce the degree of valley polarization, $S_{{\bf K}, {\bf K}^{\prime}}$, defined as
\begin{equation}
S_{{\bf K}, {\bf K}^{\prime}}=
\frac{g_{1}\left(\varphi_{\mathbf{\kappa}}\right)-g_{2}\left(\varphi_{\mathbf{\kappa}}\right)}
{g_{1}\left(\varphi_{\mathbf{\kappa}}\right)+g_{2}\left(\varphi_{\mathbf{\kappa}}\right)}.
\end{equation}
It is possible to manipulate the degree of valley polarization along a given direction by changing the polarization of the excitation. In Fig.~\ref{fig:Valley_Polar} we plot the degree of valley polarization in the instance of photocreation $S_{{\bf K}, {\bf K}^{\prime}}$ for two different orientations of the polarization plane of the excitation with respect to the crystallographic axes of a gapped graphene like material, and two different values of the band gap.

As can be seen from Fig.~\ref{fig:Valley_Polar}, a linearly polarized excitation will result in the spatial separation of carriers belonging to different valleys. This coupled with the suppression of inter-valley and back scattering should result in the accumulation of photoexcited carriers belonging to different valleys at different edges of a light spot from an illuminating laser beam. To observe this effect we propose the following experiment. Consider a gapped-2D Dirac crystal, illuminated by a laser in continuous wave operation, and under weak pumping conditions. The energy of the photoexcited carrier should be below that of a phonon which can cause intervalley scattering but large enough to cause an interband transition within the trigonal warping regime. The low-pumping regime is required to neglect carrier-carrier scattering. The linearly polarized light will create an anisotropic distribution of photoexcited carriers. These photoexcited carriers will propagate away from the light spot, with certain angles being more valley polarized than others. As they propagate away from the light spot, the photoexcited carriers will relax within their valleys in both momentum and energy, progressing towards the band edge. At the band edge, the photoexcited electrons and holes recombine to emit circularly polarized radiation with valley-dependent helicity. By fixing the laser and detector in place, as the sample is rotated the degree of circular polarization of edge-energy luminescence measured by the detector will vary with angle between the sample's  crystallographic axes and the polarization plane. This modulation will reflect the angular dependence of the degree of valley polarization induced by linearly polarized light. Notably, measuring the spatial distribution of circular polarization of edge photoluminescence to find the degree of valley polarization is somewhat similar to using optical methods to detect spin accumulation on different edges of the sample in the spin Hall effect \cite{d1971possibility,dyakonov1971current,RevModPhys.87.1213} or the optical spin Hall effect \cite{PhysRevLett.95.136601,leyder2007observation}.

\section{Optovalleytronics in the presence of the Rashba term}

Realistic graphene structures are usually grown on, or deposited atop of a substrate. The resulting asymmetry in the direction perpendicular to the graphene sheet results in an additional contribution into the spin-orbit coupling term entering the Hamiltonian. 
The modification to the spectrum can be described by adding to the pristine graphene Hamiltonian the famous Rashba spin-orbit coupling term. Following Ref.~\cite{Zarea2009}  
to account for the Rashba term, but using the $x$ and $y$ axis as in Ref.~\cite{saito1998physical} to be consistent with the previous section, yields the following Hamiltonian
\begin{equation}
    H = \begin{pmatrix}0 & \phi_0 & 0 & i\phi_{+} \\
    \bar{\phi}_0 & 0 & - i \bar{\phi}_{-} & 0\\ 
    0 & i \phi_{-} & 0 & \phi_0 \\
    - i \bar{\phi}_{+} & 0 & \bar{\phi}_0 & 0
    \end{pmatrix}\,, 
    \label{eq:HamiltonianRashba}
\end{equation}
acting on the basis $\left( \left|\psi_{\text{A}\uparrow}\right.\rangle,
    \left|\psi_{\text{B}\uparrow}\right.\rangle,
    \left|\psi_{\text{A}\downarrow}\right.\rangle,
    \left|\psi_{\text{B}\downarrow}\right.\rangle\right)^{\text{T}}$, which is an expansion of the basis used in Eq.~\eqref{eq:HamiltonianWithSOCGeneral} to include spin. Here $\bar{\phi}(k_x,k_y) = \phi(k_x,-k_y)$ and
\begin{align*}
    \phi_0(k_x,k_y) &= t e^{-i k_x 2 b/3} \left[1 + 2 e^{i k_x b} \cos \left(\dfrac{k_y a}{2}\right)\right]\,, \\
    \phi_{\pm}(k_x,k_y) &= \lambda_{\text{R}} e^{-i k_x 2 b/3} \left[1 + 2 e^{i k_x b} \cos \left(\dfrac{k_y a}{2} \pm \dfrac{2\pi}{3}\right)\right] \, ,
\end{align*}
where $b = \tfrac{\sqrt{3} a}{2}$ with $a=2.46$\AA~being the graphene lattice constant. Here $\lambda_{\text{R}}$ is the strength of Rashba spin-orbit interaction, which can be controlled by the applied back-gate voltage. Similar to Ref.~\cite{Zarea2009}, we assume $\lambda_{\text{R}} >0$.

\begin{figure}
  \centering
    \includegraphics[width=0.9\linewidth]{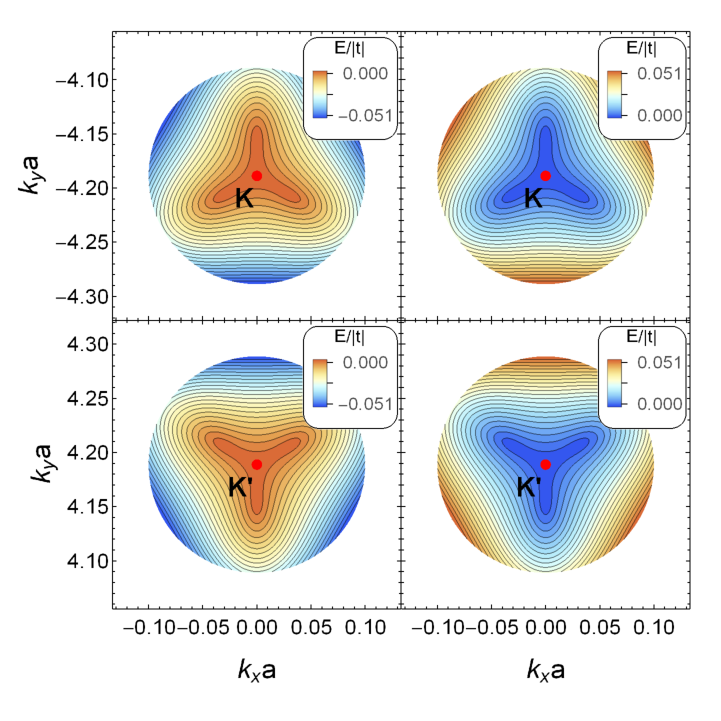}
\caption{The graphene valence and conduction energy bands around the {\bf K} and {\bf K}$^{\prime}$ points when  Rashba spin-orbit interaction is accounted for. The Rashba coupling constant is chosen to be $\lambda_{\text{R}}= 0.1 |t|$ in terms of the hopping integral, $t$. 
}
\label{fig:EnergyBandsRashba}
\end{figure}

In Fig.~\ref{fig:EnergyBandsRashba} we plot the equienergy surfaces obtained from the diagonalization of Eq.~(\ref{eq:HamiltonianRashba}) near the {\bf K} and {\bf K}$^{\prime}$ points, for carrier energies (calculated from the Dirac cone apex) three orders of magnitude smaller than the energies discussed in the previous section. 
It can be clearly seen that there is a very strong anisotropy of the equienergy surfaces of the low-energy part of the spectrum, in both the conduction and valence band of graphene.
Notably, 
the low-energy deformed triangles arising from the Rashba term are orientated in the opposite direction to the triangles due to high-energy warping discussed in the previous section. This should lead to the swap of the sign of the degree of valley polarization. 

\begin{figure}
  \centering
    \includegraphics[width=0.9\linewidth]{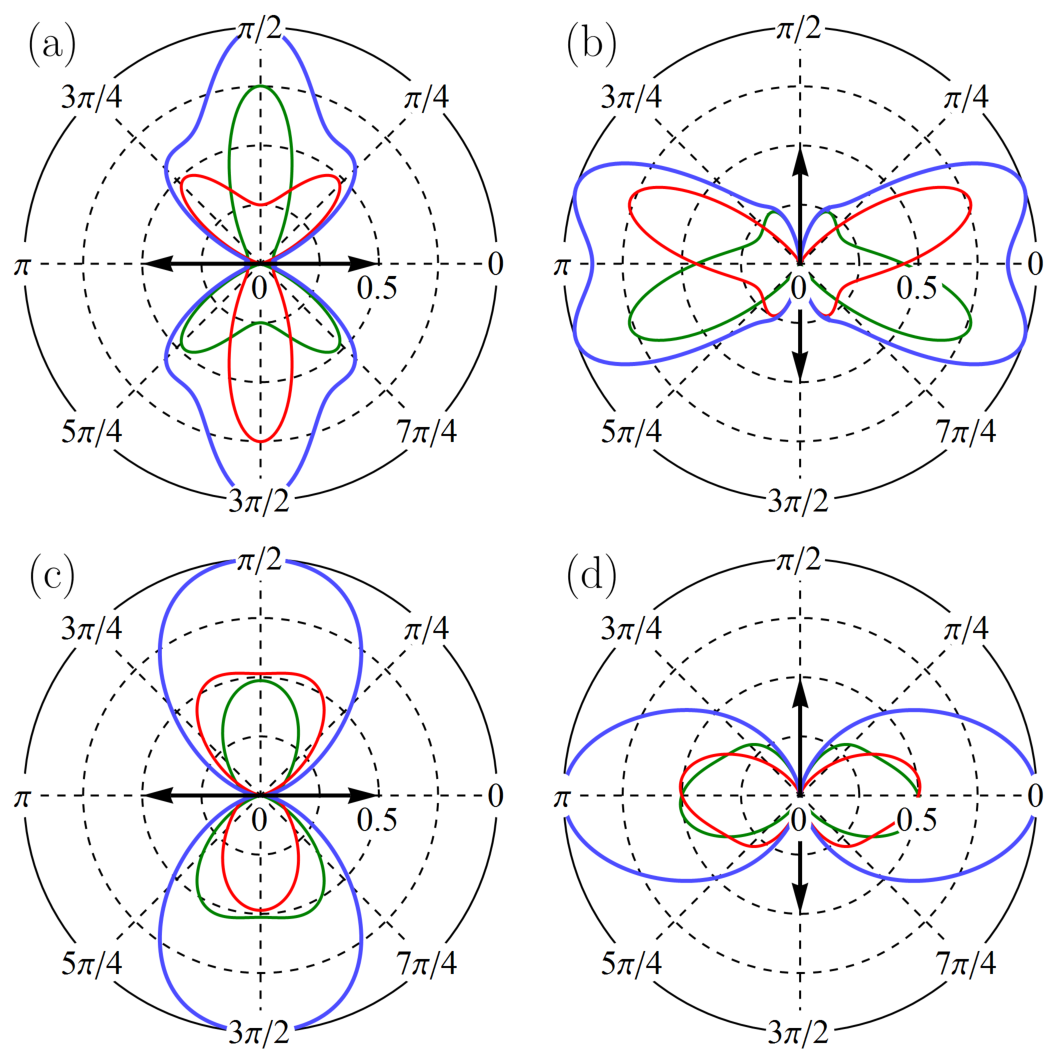}
\caption{The polar plots of the momentum distribution function of photoexcited carriers generated by linearly polarized light in graphene when Rashba spin-orbit coupling is taken into account: (a) $\phi=0$ and $h\nu= 0.05 |t|$, (b) $\phi=\pi/2$ and $h\nu= 0.05 |t|$, (c) $\phi=0$ and $h\nu = 0.3 |t|$, (d) $\phi = \pi/2$ and $h\nu = 0.3 |t|$. Here $\phi$ is the angle between the excitation polarization plane and the $x$-axis assuming normal incidence. The Rashba coupling constant is chosen to be $\lambda_{\text{R}} = 0.1\, |t|$, where $|t| \approx 3$~eV is the hopping integral in graphene. The green and red lines show the contributions from the  {\bf K} and {\bf K}$^{\prime}$ valley, respectively; while the blue contour is the total sum of valley contributions. The bold black arrows represent the polarization of the excitation.}
\label{fig:DistributionOfPhtoexcitedCarriersRashba}
\end{figure}

Differentiating the Hamiltonian given by Eq.~\eqref{eq:HamiltonianRashba} with respect to $k_x$ and $k_y$ (see Eq.~\eqref{eq:VelocityOperatorSO}), yields the $x$- and $y$-components of the velocity operators.
The momentum distribution of photoexcited carriers can be obtained by calculating the matrix elements of this velocity operator between the eigenstates of the unperturbed Hamiltonian taking into account the orthogonality of the spinor components and substituting it in Eq.~\eqref{eq:distributionfunc}, performing the integration numerically. 
We present the results of these calculations for two orthogonal linear polarizations of excitation in Fig.~\ref{fig:DistributionOfPhtoexcitedCarriersRashba}
detailing the contributions of the two valleys.
Each plot is normalized by the maximum value of the total distribution function presented by a solid blue curve. 
It can be seen from Fig.~\ref{fig:DistributionOfPhtoexcitedCarriersRashba} (a) and (b) that taking into account the Rashba term leads to the momentum alignment of photoexcited carriers for excitation energies in the far-infrared part of the spectrum. Since the degree of Rashba spin-orbit coupling can be controlled via the electric field of the back gate, it opens the door to a tunable optical valley Hall effect at infrared frequencies.

It can be seen from Fig.~\ref{fig:DistributionOfPhtoexcitedCarriersRashba} (c) and (d) that for high excitation frequencies (optical and ultraviolet) the influence of conventional trigonal warping considered in the previous section dominates. The Rashba term, which is still indeed present in this regime, provides an opposite but much smaller effect than the conventional trigonal warping. 


\section{Giant enhancement of interband transitions in narrow-gap carbon nanotubes and graphene nanoribbons}
In the first two subsections of this section we explain how the momentum alignment phenomenon in graphene leads to the giant enhancement of interband transitions in narrow-gap CNTs and GNRs. We show that in both types of quasi-one-dimensional structure the velocity matrix element at the band gap edge takes upon the universal value of $v_\mathrm{F}$. In the third subsection we discuss the potential applications of narrow-gap carbon nanotubes and graphene nanoribbons as active elements of emitters of coherent THz radiation. 

\subsection{Carbon nanotubes}
\label{subsec:tubes}
\begin{figure}
    \centering
    \includegraphics[width=0.70\textwidth]{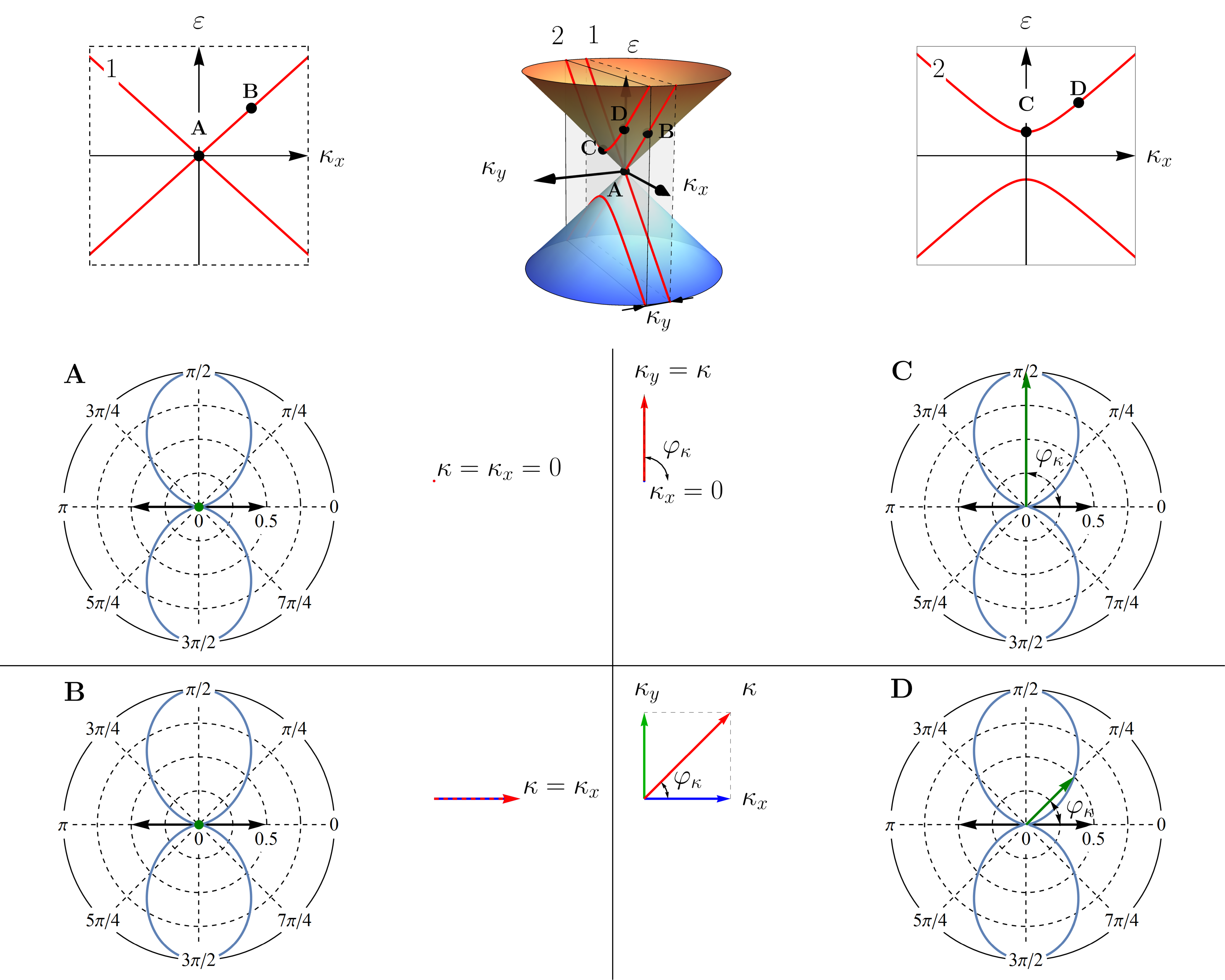}
    \caption{
     Illustration of the giant enhancement of interband transitions across the narrow band gap in quasi-metallic carbon nanotubes and graphene nanoribbons explained in terms of the momentum alignment phenomenon in graphene.
     Each of the panels A,B,C and D shows the dependence of the transition matrix element on the angle between the excitation polarization plane and the momentum of the photoexcited carrier, as well as the momentum vector direction. 
     A and B correspond to the gapless case with the energy spectrum shown in the top left-hand side panel when the cross section of the cone passes through the Dirac point ($\kappa_{y}=0$). As there is no momentum normal to the light polarization plane, optical transitions are totally forbidden in the conic approximation for both A ($\kappa_{x}=0$) and B ($\kappa_x\neq0$). C and D correspond to the narrow-gap case when the cross section avoids the Dirac point ($\kappa_{y}\neq0$). C is for a band-edge transition when $\kappa_{x}=0$ and the transition probability reaches its maximum. D corresponds to $\kappa_{y}\neq0$ and $\kappa_x\neq0$, the increase in $\kappa_{x}$ leads to a reduction in the transition probability. 
     The length of the dark-green arrows in the polar plots (vanishing for A and B) show the magnitude of the matrix element of transition. }
    \label{fig:Explaination}
\end{figure}

In the absence of curvature effects, narrow-gap CNTs can be considered as an unrolled graphene sheet, subject to the periodic boundary condition: $k_{y} \cdot {C}_h =  2 \pi \ell$, where ${k}_{y}$ is the electron wave vector (i.e. momentum) quantized along the circumferential direction chosen to be along the $y$-axis, ${C}_h$ is the tube circumference and $\ell$ is an integer. The different subbands in the 1D reciprocal space of the nanotube form a set of parallel cutting lines in the 2D reciprocal space of the graphene sheet~\cite{Samsonidze2003}. If $n-m = 3p$, where $p$ is an integer (see Ref.~\cite{saito1998physical} for classification of tubes), then the cutting line crosses the graphene Dirac point exactly. In this instance, the effective matrix Hamiltonian of the tube can be written in the same form of the graphene Hamiltonian, Eq.~\eqref{eq:HamiltonianWithSOC}, with $\Delta=\kappa_{y}=0$ and $\hbar \kappa_{x}$ playing the role of the free electron momentum along the nanotube axis measured relative to the Dirac point of interest. As $\kappa_{y}$ is zero, $\varphi_{\kappa}$ is always zero for any $\kappa_x$ (see its definition after Eq.~\eqref{eq:EigenvectorsWithSOC1}). Then, since the polarization of light is directed along the tube ($x$-axis) (i.e. in Eqs.~\eqref{eq:DistributionFunctionOfPhotoExcitedCarriers} and~\eqref{eq:trans_prob} $\phi$ is zero), the corresponding angular generation density is zero as shown in diagrams for points $\mathbf{A}$ and $\mathbf{B}$ in Fig.~\ref{fig:Explaination}. However, due to curvature effects, the aforementioned CNTs possess a narrow-gap~\cite{kane1997size,Zhou2000,Ouyang2001,Shyu2002,Dyachkov2018}, $\varepsilon_{g}$, which is of the order of THz. The effective Hamiltonian of the tube can still be written in the same form of an effective graphene sheet, with $\Delta$ remaining zero in Eq.~\eqref{eq:HamiltonianWithSOC}, but $\kappa_{y}$ being finite due to band gap fixed by the curvature  or manipulated by a magnetic field. We therefore set $2\hbar v_\mathrm{F}\kappa_{y}=\varepsilon_{g}$ in Eq.~\eqref{eq:HamiltonianWithSOC}.  For finite $\kappa_{y}$, $\varphi_{\kappa}$ can take a range of values, and is precisely $\pi/2$ at the band gap edge. This is illustrated by diagrams for points $\mathbf{D}$ and $\mathbf{C}$ in Fig.~\ref{fig:Explaination}. Substituting $\varphi_{\kappa}=\pi/2$ into Eq.~\eqref{eq:DistributionFunctionOfPhotoExcitedCarriers} results in the absolute value of the matrix element of velocity being precisely $v_{F}$. Therefore, any small gap, irrespective of its origin, leads to a strong spike-like dependence of the velocity matrix element on electron wave vector in the vicinity of the Dirac point with a peak value equal to the Fermi velocity in graphene.

\begin{figure}[htb]%
\centering
\includegraphics*[width=\textwidth]{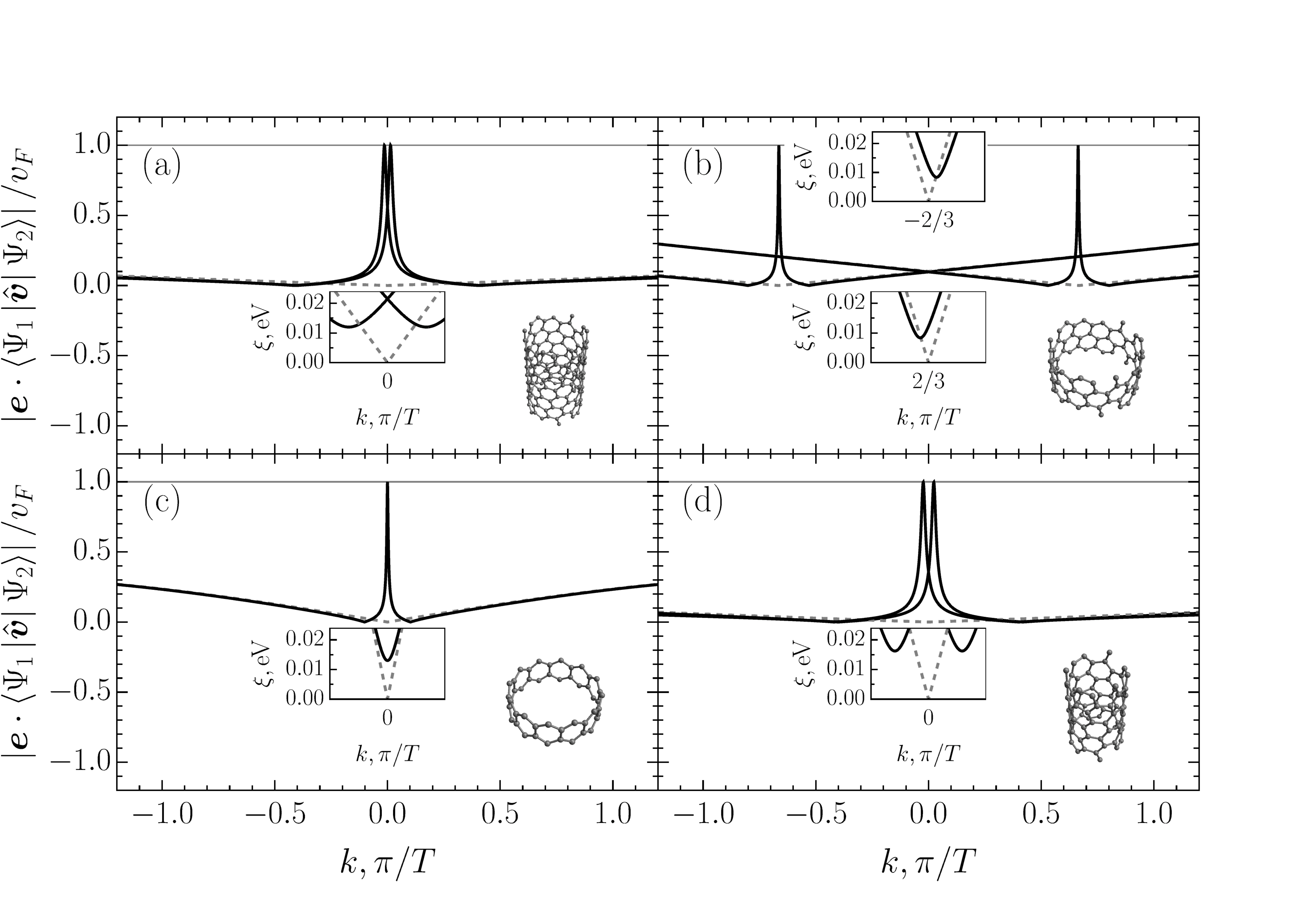}
\caption{The absolute value of the velocity operator matrix element with (solid black) and without (dashed grey) the curvature effect for (a) CNT $(9,3)$; (b) CNT $(12,3)$; (c) CNT $(12,0)$ and (d) CNT $(6,3)$. The insets show the conduction bands in the vicinity of the Dirac point with (solid black) and without  (dashed grey) the curvature effect taken into account. The CNT unit cells are presented in the right bottom corner of each plot. For all chosen tubes, only the carbon-carbon bond contraction which dominates the curvature effect is accounted for. The universal character of the peak is highlighted by the solid horizontal line which marks $v_F$ level. $T$ is the translation period of the tube.}
\label{fig:CNTvmes}
\end{figure}

In Fig.~\ref{fig:CNTvmes} we plot the absolute value of the matrix element $\boldsymbol{e}\cdot\left\langle \Psi_{1}\left|\hat{\boldsymbol{v}}\right|\Psi_{2}\right\rangle$ as a function of the electron wave vector, $k$, in the whole CNT Brillouin zone. 
The shape of the presented peaks are described by the second term in the expression:
\begin{equation}
v_{F}  \left( \frac{a_0 \cos (3 \vartheta)}{4}  \kappa_{x} - \dfrac{\kappa_{y} }
{\sqrt{ \kappa_{x} ^2 +  \kappa_{y} ^2}} \right) \, ,
\label{eq:CNTvme}
\end{equation}
where $\kappa_{x} $ is the absolute value of the momentum measured relative to the band gap edge, $\kappa_{y} = \varepsilon_{ g}/( 2 \hbar v_F)$, with $\varepsilon_{g}$ being the band gap and $a_0 = 0.142$ nm is the distance between nearest carbon atoms. The angle $\vartheta$ is the chiral angle which has a value of $0$ for zigzag and $\pi/6$ for armchair tubes. The $\kappa_{y}$ represents the shift of the cutting line~\cite{Samsonidze2003} away from the Dirac point, whereas the cutting lines themselves are equidistantly separated in momentum space by $k_y = 2 \pi \ell/C_{h}$, where $C_h$ is the tube circumference. In other words, the separation is given by the size of the electron momentum quantized along the circumferential direction. Within this formalism one can also incorporate the effect of an external magnetic field. A magnetic field applied along the tube axis shifts cutting lines altogether, thereby increasing $\kappa_{y}$ for the lines passing through the Dirac points. For an armchair nanotube this results in the opening of the gap, accompanied by strongly allowed interband transitions at the band edge~\cite{Portnoi2009,Moradian2010,Chegel2014}. For quasi-metallic tubes the shift in the cutting lines modifies the band gap and lifts the degeneracy~\cite{Hartmann_IOP_2015}.

It must be pointed out that whether it be strain or applied magnetic field, the quantity that matters for the enhanced absorption or emission across the band gap is the shift of the cutting line away from the Dirac point. The effects of curvature and magnetic field can be both incorporated into the model in question:
\begin{equation}
\kappa_{y} = \kappa_{y,s} + \kappa_{y,\Phi} \, .
\end{equation}
Here $\kappa_{y,s} $ is the shift away from the {\bf K} point due to intrinsic strain and $\kappa_{y,\Phi}$ is the shift of the cutting line due to a magnetic field applied along the tube axis. The contribution to $\kappa_{y,s}$ due to C-C bond length contraction can be written in terms of the hopping integrals, $t_i$, that are now different for three nearest-neighbors:
\begin{equation}
\kappa_{y,s} =  \frac{2 \sin(\vartheta) \tau_2 - \left[\sqrt{3} \cos(\vartheta)- \sin(\vartheta)\right] \tau_3}{ a (t_1 + t_2 + t_3)}  \,
\end{equation}
where $a=\sqrt{3} a_0$ is the graphene lattice translation constant, $\tau_2 = (t_1 - 2 t_2 + t_3)$ and $\tau_3 = (t_1 + t_2 - 2 t_3)$. The shift due to the applied magnetic field is given by the expression:
\begin{equation}
\kappa_{y,\Phi} = \dfrac{2 \pi }{C_h} \, \dfrac{\Phi_{\phantom{0}}}{ \Phi_0} \, ,
\end{equation}
where $\Phi$ and $\Phi_0$ are the magnetic flux through the tube cross-section and the magnetic flux quantum, respectively.

Equation~\eqref{eq:CNTvme} is distinctly different to the velocity matrix element obtained in Ref.~\cite{Kibis2007}, which contains only the term linear in $\kappa_{x}$ and completely overlooks the term $\kappa_{y}/\sqrt{ \kappa_{x} ^2 +  \kappa_{y} ^2}$, which is associated with the angular dependence of the transition matrix element in graphene discussed in Sec.~\ref{sec:Rules}. For direct transitions associated with THz frequency excitations, the second term appearing in Eq.~\eqref{eq:CNTvme} completely dominates, and therefore the scheme put forth in Ref.~\cite{Kibis2007} neglects the most important contribution to the matrix element of velocity. The paper in question~\cite{Kibis2007} proposes a scheme based on the electric-field-induced heating of an electron gas, resulting in the inversion of population of optically active states with energy difference within the THz spectral range~\cite{Kibis2007}.  It implies that for certain types of carbon nanotubes the heating of electrons to energies below the high-energy phonon emission threshold results in spontaneous THz emission with peak frequency controlled by an applied voltage. The spectral density of spontaneous emission was found to be proportional to $\nu^3$, where $\nu$ is the photon frequency. This scheme, however, overlooks two important issues: firstly, the presence of the curvature induced band gap, which is of the order of THz, and secondly, the dependence of the transition matrix element on the angle between the light polarization vector and the effective ``total'' momentum including the fictitious curvature-induced component. Correctly taking into account these effects results in the spectral density of spontaneous emission not being heavily suppressed near the Dirac point. A pronounced peak at the band edge exists, which is highly desirable for THz frequency range applications~\cite{HartmannRev2014}. Only at larger optical excitation energies does the spontaneous emission become proportional to the cube of frequency.

\subsection{Graphene nanoribbons}
\label{subsec:ribbon}
Let us now compare carbon nanotubes with similar graphene nanoribbons. The periodic boundary condition applied to the tube, must be replaced with the so-called ``hard wall" boundary condition~\cite{Saroka2017}. In the absence of edge effects, the band structure of an armchair GNRs (AGNRs) can be obtained from the band structure of graphene using a cutting line technique similar to that used for CNTs. For armchair GNRs, transverse electron momentum is a good quantum number, therefore the hard wall boundary condition leads to ${k}_y\cdot {L} =  \pi \ell$, where $L$ is the ribbon width and $k_{y}$ is the electron transverse momentum chosen to be directed along the $y$-axis. These two types of boundary conditions match if $L= C_h / 2$. This occurs, for example in AGNRs$(N)$, where $N$ denotes the number of C atom pairs in the unit cell of the ribbon, and zigzag CNTs$(N+1,0)$~\cite{White2007,Saroka2018}. For these specifically chosen structures the electronic properties are almost identical. At low energies, the band spectra of these tubes duplicate those of the ribbons. The only crucial difference is that many tubes bands are double-degenerate, whereas corresponding ribbons bands are not. At higher energies the band structures deviate from one another. Namely, the spectrum of tubes contains some energy bands which are absent in the ribbons. The edge-effect in armchair ribbons can be incorporated into the tight-binding model as corrections to the hopping integrals at the ribbon edges~\cite{Zheng2007}. 
Similarly to the curvature effect for nanotubes, the influence of the edge effect on the optical properties of nanoribbons can be described by introducing a fictitious momentum normal to the ribbon's free-motion axis. Then the velocity matrix element for the transition between the highest valence subband and the lowest conduction subband in a narrow-gap GNR is of a similar form to Eq.~\ref{eq:CNTvme}, with $\kappa_y$ now related to the GNR band gap. Thus, the momentum alignment phenomenon again leads to the giant enhancement of the band-edge transition rate.

\begin{figure}
	\includegraphics*[width=0.85\textwidth]{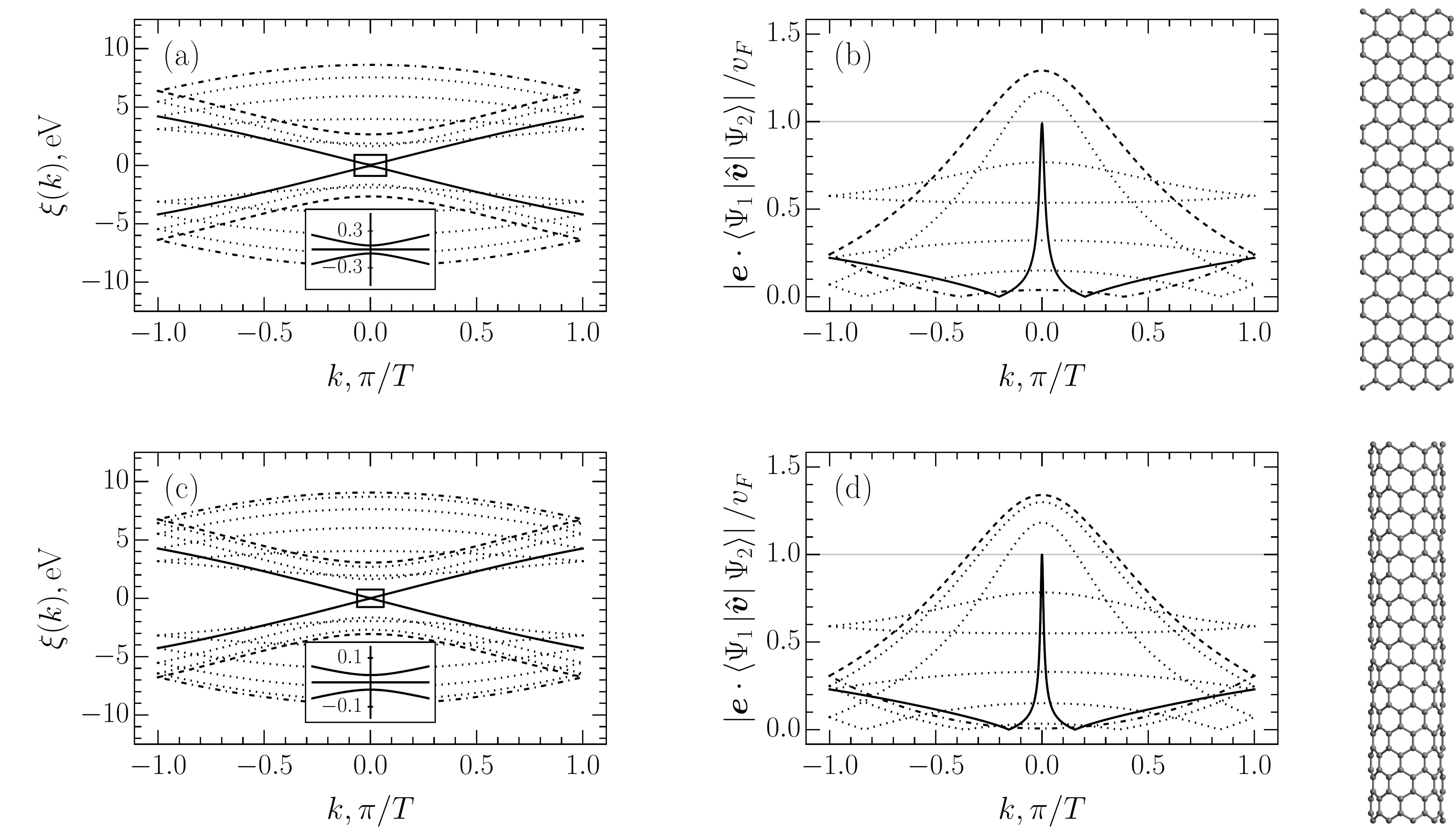}
	\caption{
(a),(c) are the  band structures and  (b),(d) the velocity operator matrix elements of a AGNR$(8)$ and zigzag CNT$(9,0)$, respectively.  Transitions between the closest valence and conduction subbands (thick black line), the lowest and highest subbands (dashed dotted, light gray line), and for the subbands, for which velocity matrix element attains the maximum possible value (dashed, gray line), are highlighted with respect to the remaining bands and matrix elements (gray, dotted-line). The insets in panels (a) and (c) show the zoomed in region close to the Dirac point where the band gap is present. In panels (b) and (d) the solid horizontal line corresponds to $v_\mathrm{F}$ as a guide to the eye. On the right side the atomic structures are shown. In both cases the hopping integral, $|t| \approx 3$~eV and the edge correction for the ribbon is $0.05 \, |t|$, whereas the curvature correction for the tube is $0.01 \, |t|$. $T$ is the translation period of the structure.
	}
	\label{fig:AGNRvsZCNT}
\end{figure}

In Fig.~\ref{fig:AGNRvsZCNT} (a), (c) we show the electronic band structure of a AGNR and zigzag CNT for the case of $N=8$, taking into account the edge effect in the ribbon and the curvature effect in the tube. It can be seen from the figure that the degree of equivalence for the two energy band structures is incredibly high throughout the whole Brillouin zone. In Fig.~\ref{fig:AGNRvsZCNT} (b), (d), we demonstrate that the equivalence in the electronic properties extends to the optical transition selection rules in the presence of the curvature effect for zigzag CNTs and the edge effect for the AGNRs. Similar to the energy bands, the majority of curves plotted in~Fig.~\ref{fig:AGNRvsZCNT} (d) for the velocity matrix element of the CNT$(9,0)$, are twice degenerate compared to the curves presented in~Fig.~\ref{fig:AGNRvsZCNT} (d) for the AGNR$(8)$. The velocity matrix element plot for the CNT contains some curves which are absent in~Fig.~\ref{fig:AGNRvsZCNT} (b) for the AGNR. These curves correspond to transitions between energy bands that are present only in the tube and are absent in the energy spectrum of the ribbon. The velocity matrix element for transitions between the lowest conduction and highest valence subbands that shows a giant enhancement due to the intrinsic strain effects are depicted by the thick black curves. As one can see, the peaks related to the enhancement are very similar in tubes and ribbons. In both structures these peaks have a maximum magnitude of $v_F$, and the only distinguishable difference between them is their widths. The AGNR peak is wider because the value of the hopping integral edge correction is larger than the correction for the tube due to the curvature effect. As in the case of Fig.~\ref{fig:CNTvmes}, only the shortening of the C-C bond is taken into account in~Fig.~\ref{fig:AGNRvsZCNT} (d).

It should be noted that a similar transition enhancement may be found in a chemically functionalized graphene sheets lined with periodic patterns of hydrogen, oxygen or fluorine adatoms to form 2D superlattices with 1D graphene channels. Indeed, the electronic properties of the resulting graphene channels are known to be similar to those of GNRs~\cite{Chernozatonskii2007}. The edge-like effects in the channels arise from distortions caused by sp$^3$ hybridization of carbon atoms, which form chemical bonds with adatoms. The low-energy electronic properties are still determined by $\pi$-orbitals; therefore, the theoretical treatment of their optical properties should be essentially the same as for nanoribbons. An additional tunability of the absorption frequencies of different channels can be achieved by stretching the whole graphene sheet~\cite{Chernozatonskii2010}.

\subsection{Utilizing enhanced transitions in carbon nanotubes and graphene nanoribbons}
The interband transition probability rate per unit volume is proportional to the product $\rho\left|\boldsymbol{e}\cdot\left\langle \Psi_{1}\left|\hat{\boldsymbol{v}}\right|\Psi_{2}\right\rangle \right|^{2}$, where $\rho$ is the joint density of states, which unlike in 2D systems, diverges at the band gap edge for a 1D system exhibiting a Van Hove singularity. This divergence in the density of states means that 1D systems effectively behave like quantum dots. At the band gap edge the matrix element of velocity has a value of $v_F \approx 10^6$~m/s. This coupled with the presence of the Van Hove singularity gives rise to a large interband transition probability rate which should make the experimental detection of these transitions possible. In principle, these transitions can be detected either by absorption or emission  spectroscopy methods. The major difficulty for both types of experiments is related to the size of the band gap corresponding to THz frequencies. On one hand this makes these transitions difficult to detect by conventional spectroscopy techniques, on the other hand it makes them ideal for THz applications~\cite{Hartmann2019}.

Let us first consider absorption experiments. In these types of measurements one usually faces two principal difficulties. Firstly, real structures are always spuriously doped by the chemicals used in the sample preparation.  For example, the bottom of the lowest conduction subband in CNTs can be filled with electrons up to $\sim 120$~meV~\cite{Kampfrath2008}, which means the low-energy transitions across the band gap are suppressed due to Pauli blocking. Secondly, there is another competing and notoriously strong mechanism - the absorption due to free carriers, i.e. plasmonic absorption. It has been demonstrated that the broad absorption peak in the region of $\sim 1-2$ THz for CNTs~\cite{Ugawa1999} is of plasmonic origin~\cite{Slepyan2010,Shuba2012,Ren2013}. This peak falls within the same frequency range associated with interband transitions thus making their detection during absorption measurements difficult.

\begin{figure}[t]%
	\includegraphics*[width=0.8 \linewidth]{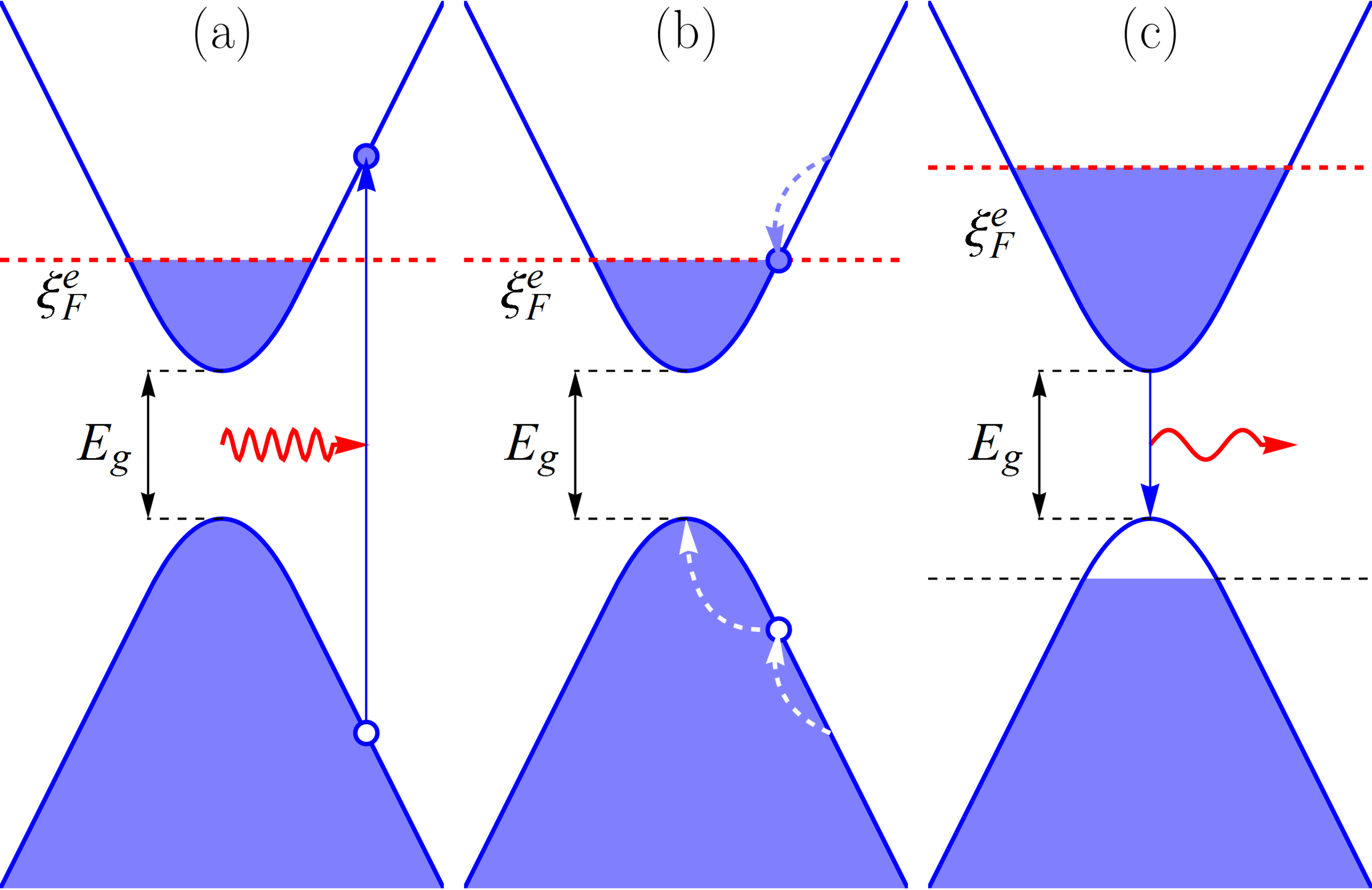}
	\caption{%
		A schematic illustration of (a) the high frequency optical excitation (b) non-radiative electron relaxation due to the electron-phonon scattering and (c) the population inversion in an $n$-doped narrow gap CNT or GNR.}
	\label{fig:Excitation_scheme}
\end{figure}

However, these transitions could be detected by measuring THz emission from the sample pumped at optical frequencies. As schematically depicted in Fig.~\ref{fig:Excitation_scheme} (a), a valence subband electron can be excited above the Fermi level into the conduction subband. 
Illuminating the sample using a broad range of optical frequencies will lead to the promotion of many electrons into the conduction band, and the creation of holes in the valence band. 
The photoexcited electrons and holes quickly thermalize with the lattice due to the scattering on phonons ($\tau_{ph} \sim 3$~ps~\cite{Park2004}). As a result of this process the holes float up to the top of the valence band, creating a population inversion, whereas the excited electrons join the Fermi sea contributing to the increase of the non-equilibrium quasi Fermi level in the conduction band. The final stage of the process, the emission of photons of the band-gap frequency, is sketched in the Fig.~\ref{fig:Excitation_scheme} (c). This transition occurs with an extremely high probability, since the optical matrix element is maximal at the band gap edge and the density of states diverges. These transitions should be easily distinguished from the thermal emission of a laser-heated array of aligned CNTs or GNRs, since the interband THz emission is very narrow in frequency and strongly polarized along the quasi-1D nanostructure axis. It should be noted that the absorption by free carriers in the sample can be minimized by the proper choice of structure length, since the plasmonic resonance is a geometrical one and depends strongly on the structures longitudinal size~\cite{Slepyan2010,Shuba2012}.



One advantage tubes and ribbons have for THz applications is that their band gap can be manipulated by external fields, thus providing tunability of the emission frequency. For tubes this can be achieved by applying a magnetic field along the tube's axis, while for ribbons an electric field can be applied in-plane~\cite{Chang2006,Saroka2015}. This tunability in the transition frequency may be employed as a criteria for distinguishing the THz radiation generated by interband transitions from that resulting from the photocurrent in semiconducting structures~\cite{Titova2015}.

The results presented in this paper were based on a single electron picture. Excitonic effects are known to dominate the optical properties of semiconducting structures~\cite{KonoShaver07,Denk2014}. However, in narrow gap CNTs and AGNRs the exciton binding energy has been shown to never exceed the bandgap~\cite{Hartmann2011,PhysRevA.95.062110}. Therefore, at room temperature the electron-hole pairs should be fully ionized. Hence, undesirable effects due to dark excitons~\cite{Srivastava2008} should not dominate optical processes in narrow-gap nanotubes. It should be also mentioned that the emission output can be maximized by putting narrow gap quasi-metallic CNT or GNR samples into a microcavity similar to what has been done for semiconducting CNTs~\cite{He2018,Gao2018}. THz mirrors with low losses should be carefully designed~\cite{Headland2017} to achieve gain in this case. The analysis of losses in such systems is a subject of investigation in future research.

Synthesis techniques for suitable structures are developing at a fast pace, for example, AGNRs of the metallic family can already be produced with atomically smooth edges~\cite{Zhang2015}, and there is also rapid progress towards the production of highly aligned arrays of carbon nanotubes with selected chirality~\cite{Gao2019}. 
This progress allows us to envisage that the effect of giant enhancement of interband transitons in narrow-gap quasi-1D nanostructures, discussed in this paper, can be utilized in tunable sources of coherent THz radiation in the near future.


\section{Conclusions}

We have calculated the momentum distribution function of photoexcited carriers created by linearly polarized light in 2D Dirac materials. The momentum distribution function is highly anisotropic and the alignment is dictated by the orientation of the polarization vector. It has been shown that in the high energy regime, i.e. in the presence of trigonal warping, one must take into account the contributions from both \textbf{K} points to achieve a physically meaningful result. This warping has also been shown to lead to the spatial separation of photoexcited carriers belonging to two different valleys (optical valley Hall effect), therefore opening the door to the optical control of valley polarization (optovalleytronics) in 2D Dirac materials.
Notably, the best-known trigonal warping effects become important at high energies far away from the apex of the Dirac cone, making light-induced valley polarization possible at optical frequencies, with the effect becoming stronger with increasing photon energy. 

In gapped 2D Dirac materials, the optical control of valley polarization can also be achieved via the well-known effect of using circularly polarized light. In these materials, the optical selection rules associated with linearly polarized light of near-band-gap energies are valley-independent, in stark contrast to the valley-dependent optical selection rules associated with circularly polarized light. This valley dependence of the circularly polarized transitions can be utilized to measure the degree of valley polarization induced by linearly polarized light of high (well above the band gap) energies, by analyzing the degree of circular polarization of the band edge luminescence at different sides of the light spot. 

In a gapless Dirac material, there is also a more subtle trigonal warping effect very close to the Dirac point, caused by the Rashba spin-orbit term, which results in the splitting of each Dirac cone into three closely separated mini-cones. Due to this effect valley polarization can also be induced by linearly polarized light in the far infrared regime. The strength of the Rashba splitting and the relevant energy scale can be conveniently adjusted via a back gate, thus providing an additional control over the strength of the optical valley Hall effect.



The momentum alignment phenomenon in graphene, which is the focus of this paper, allows a simple physical explanation of the giant enhancement and universality of the low-energy interband transitions in narrow-gap carbon nanotubes and armchair graphene nanoribbons. In the total absence of a band gap in these quasi-1D systems, optical transitions induced by light polarized along their axes are strictly forbidden in the conical approximation. Indeed, the selection rules in graphene dictate that the momentum of photoexcited carriers should align normal to the light polarization plane.
The opening of the gap, either by curvature or edge effects, introduces an effective momentum normal to the structure's axis resulting in the universal value of the velocity matrix element with an amplitude equal to the Fermi velocity in graphene. Moving away from the band gap edge leads to a drastic reduction in the transition matrix element until it starts to increase with increasing frequency due to the trigonal warping effect.
The peak in the matrix element of velocity, which coincides with the van Hove singularity at the band gap edge, makes the considered quasi-1D systems promising candidates for active elements in coherent THz radiation emitters.

\section*{Acknowledgements}
This work was supported by the EU FP7 ITN NOTEDEV (FP7-607521), EU H2020 RISE projects CoExAN (H2020-644076), TERASSE (H2020-823878) and DiSeTCom (H2020-823728). R.R.H. acknowledges financial support from URCO (14 F 1TAY20-1TAY21). 

This article was published in the special issue of Journal of Experimental and Theoretical Physics dedicated to the 95th birthday of E.I. Rashba.

%

\end{document}